\documentclass{elsarticle}
\usepackage{amsmath}
\usepackage{latexsym}
\usepackage{amsfonts}
\usepackage{graphicx}
\usepackage{mathrsfs}
\usepackage{hyperref}
\usepackage{mathrsfs}
\usepackage{accents}
\usepackage{xcolor}
\usepackage{color}

\begin{document}

\title{Constraining $f(R)$ gravity in solar system, cosmology and binary pulsar systems}

\author[ustc,as]{Tan Liu \corref{cor}}
\cortext[cor]{Corresponding author}
\ead{lewton@mail.ustc.edu.cn}

\author[ustc,as]{Xing Zhang}
\ead{starzhx@mail.ustc.edu.cn}

\author[ustc,as]{Wen Zhao}
\ead{wzhao7@ustc.edu.cn}

\address[ustc]{CAS Key Laboratory for Researches in Galaxies and Cosmology, Department of Astronomy, University of Science and Technology of China, Chinese Academy of Sciences, Hefei, Anhui 230026, China }
\address[as]{School of Astronomy and Space Science, University of Science and Technology of China, Hefei 230026, China}
\begin{abstract}
The $f(R)$ gravity can be cast into the form of a scalar-tensor theory, and scalar degree of freedom can be suppressed in high-density regions by the chameleon mechanism. In this article, for the general $f(R)$ gravity, using a scalar-tensor representation with the chameleon mechanism, we calculate the parametrized post-Newtonian parameters $\gamma$ and $\beta$, the effective gravitational constant $G_\text{eff}$, and the effective cosmological constant $\Lambda_\text{eff}$. In addition, for the general $f(R)$ gravity, we also calculate the rate of orbital period decay  of the binary system due to gravitational radiation. Then we apply these results to specific $f(R)$ models (Hu-Sawicki model, Tsujikawa model and Starobinsky model) and derive the constraints on the model parameters by combining the observations in solar system, cosmological scales and the binary systems.
\end{abstract}



\maketitle

\section{Introduction}

Since the discovery of cosmic acceleration in 1998 \cite{1538-3881-116-3-1009,0004-637X-517-2-565}, considerable efforts have been devoted in cosmology to understand the physical mechanism responsible for it. The $\Lambda$CDM model interprets the acceleration of the universe as a consequence of the cosmological constant. Although this model matches cosmological observations well \cite{ade2016planck}, the cosmological constant suffers from some theoretical problems. If the cosmological constant originates from the vacuum energy in quantum field theory, extreme fine-tuning is required to explain its smallness \cite{RevModPhys.61.1}.  It is also difficult to explain its closeness to the present matter density of the universe \cite{RevModPhys.61.1}.
This motivates the search for alternative explanations for the cosmic acceleration.

Two types of approaches have been considered. One can either introduce a new kind
of matter whose role is to trigger acceleration, or modify the
behavior of gravity on cosmological scales \cite{0253-6102-56-3-24,Joyce20151}. In the first
approach, dark energy is introduced as a new energy form, which has positive energy density but negative pressure. In the second approach, various attempts to
modify gravity have been presented. For  recent reviews on modified gravity, see \cite{CLIFTON20121,NOJIRI20171,burrage2017tests,ANDP:ANDP201400058}.

Lovelock's theorem states that General Relativity (GR) represents the most general theory describing a single metric that in four dimensions has field equations with at most second-order derivatives \cite{lovelock1971einstein}. As a result of this theorem, one way to modify Einstein's field equations is to permit the field equations to be higher than second order.
In this paper, we will consider the so-called $f(R)$ gravity  which has  fourth order field equations. The Ricci scalar $R$ in the gravity action is replaced by  a general function of Ricci scalar $f(R)$. For reviews on $f(R)$ gravity, see \cite{RevModPhys.82.451,de2010f}. The $f(R)$ gravity does not introduce any new type of matter and can lead to the late time acceleration of the universe\cite{Xu2014,starobinsky2007disappearing}. When cast into the scalar-tensor theory, the $f(R)$ gravity implies a strong coupling between the scalar field and matter. This would violate all experimental constraints on deviations from Newton's gravitation \cite{Brax2008}.  Certain constraints have to be imposed on the function $f(R)$ for the model to be linearly stable \cite{PhysRevD.74.104017,PhysRevD.72.124005} and pass local gravitational tests \cite{PhysRevD.77.023507}. The first attempt $f(R)=R-\mu^{2(n+1)}/ R^n$ proposed by Carroll \textit{et al.} in \cite{PhysRevD.70.043528} failed these constraints right away. However, since then, models that evade them have been found \cite{PhysRevD.68.123512,NOJIRI2007343}. Fortunately, the chameleon mechanism can alleviate these constraints. Imposing the chameleon mechanism, the scalar field can develop an environment dependent mass \cite{Brax2017,PhysRevLett.93.171104,PhysRevD.69.044026}. When  the ambient matter density is large enough, its mass becomes large, and the corresponding  fifth force range is short. Thus the scalar field can be hidden in the high density environment and the fifth force cannot be detected \cite{Brax2008}.

The parametrized post-Newtonian (PPN) formalism is useful to study different theories of gravity \cite{Will2014,will1993theory,misner1973gravitation,weinberg1972gravitation}. In the PPN formalism, the PN (weak field and slow motion) limit of different theories are characterized by a set of PPN parameters and the most important two parameters are $\gamma$ and $\beta$. These two parameters can be directly measured by the solar system experiments. The GR prediction  ($\gamma=1$ and $\beta=1$) is consistent with the observations \cite{bertotti2003test}, which provide  constraints on various modified gravity models \cite{PhysRevD.72.044022,PhysRevD.72.083505}. Meanwhile, the binary pulsar systems can emit gravitational waves and provide a good test for gravitational theories \cite{Will2014,will1993theory,neutron,hou2017constraints,Stairs2003}. Since these systems lose energy due to gravitational radiation, the orbital period of these systems will decay \footnote{The LIGO Scientific Collaboration and Virgo Collaboration have detected the gravitational waves \cite{PhysRevLett.116.061102,PhysRevLett.116.241103,PhysRevLett.118.221101,Abbott2017a,Abbott2017}. This is an important milestone and opens new windows in the gravitational physics and astrophysics.}.
Several authors have considered this effect in $f(R)$ gravity \cite{dyadina2016verification,de2013testing,de2015probing} for some specific models.
However, in these works, the authors have ignored the chameleon mechanism . Although some authors have applied
the chameleon mechanism to $f(R)$ gravity when they study the PN limit, they only calculate the PPN parameter $\gamma$ \cite{PhysRevD.76.063505,Hu:2007nk}.

In this paper, we give a comprehensive investigation on various constraints on the general $f(R)$ gravity with chameleon mechanism. Following the method developed in our previous work \cite{Zhang2016}, we first calculate the PPN parameters $\gamma$ and $\beta$, the effective cosmological constant, and the effective gravitational constant in the general $f(R)$ gravity. Considering the current observations in solar system and cosmological scales, we derive the combined constraint for the general $f(R)$ gravity. Binary pulsar system is a good testing ground for alternative theories of gravity. In the previous work \cite{Zhang2017}, we have derived the orbital period derivative for quasicircular binary systems in scalar-tensor gravity with chameleon mechanism. Here, applying the similar analysis to $f(R)$ gravity, we obtain the orbital period derivative for quasicircular  neutron star-white dwarf (NS-WD) binary systems. Using the observational data of PSR J0348 +0432 \cite{Antoniadis1233232} and PSR J1738 +0333 \cite{freire2012relativistic}, we also obtain the binary pulsar constraints on $f(R)$ gravity. We find that the chameleon mechanism cannot apply to Palatini $f(R)$ gravity. Thus, in the paper, we mainly focus on metric $f(R)$ gravity. Applying the general results to the specific $f(R)$ models, including Starobinsky model, Hu-Sawicki model and Tsujikawa model, we obtain the constraints on the model parameters.

The paper is organized as follows: In Sec. \ref{fr_cha}, we review $f(R)$ gravity and chameleon mechanism. In
Sec. \ref{constraint}, we study various observational constraints on $f(R)$ gravity, and obtain the parameter constraints on the specific models. We conclude in Sec. \ref{conclusion}.

Through out this paper, the metric convention is chosen as $(-,+,+,+)$, and Greek indices $(\mu,\nu,\cdots)$ run over $0,1,2,3$. We set the units to $c=\hbar=1$, and therefore the reduced Planck mass is $M_\text{Pl}=\sqrt{1/8\pi G}$, where $G$ is the gravitational constant.

\section{$f(R)$ gravity and Chameleon mechanism}\label{fr_cha}

The $f(R)$ gravity comes about by a straightforward generalization of the Ricci scalar $R$ to become a general function  $f(R)$ in the action for gravity. When varying the action, there exist two formalisms: the metric formalism and the Palatini formalism. In the Palatini formalism, the connection is not taken to be the Levi-Civita connection of the metric \textit{a priori} and one varies the action assuming that the metric and the connection are independent variables. Although these two formalisms lead to the same field equations in GR \cite{wald1984general}, this is no longer true for $f(R)$ gravity. We will investigate these two formalisms respectively.

\subsection{Metric $f(R)$ gravity}
The total action for metric $f(R)$ gravity takes the form \cite{RevModPhys.82.451}
\begin{equation}
\label{fr}
S={1\over 16\pi G}\int d^4 x\sqrt{-g}\,f(R)+ S_m(g_{\mu\nu},\Psi_m),
\end{equation}
where $\Psi_m$ denotes all the matter fields.
Variation with respect to the metric $g_{\mu\nu}$ gives the field equations \cite{RevModPhys.82.451}
\begin{equation}
f'(R)R_{\mu\nu}-\frac12 f(R)g_{\mu\nu}-[\nabla_\mu\nabla_\nu-g_{\mu\nu}\square]f'(R)=8\pi G T_{\mu\nu},
\end{equation}
where a prime denotes differentiation
with respect to $R$ and $\square=\nabla^\mu\nabla_\mu$. Since the field equations contain the second derivative of $R$ and  $R$ includes second derivatives of the metric, the field equations are fourth order partial differential equations in the metric.

Handling  fourth order equations can be troublesome, but $f(R)$ gravity can be recast as a scalar-tensor theory via a conformal transformation and the corresponding field equations become second order. Conformal transformation of the metric can also show the scalar degree of freedom explicitly.
Introducing a new field $\chi$, we obtain a dynamical equivalent action \cite{RevModPhys.82.451}
\begin{equation}
\label{equiv}
S={1\over 16\pi G}\int d^4 x\sqrt{-g}\,[f(\chi)+f'(\chi)(R-\chi)]+ S_m(g_{\mu\nu},\Psi_m).
\end{equation}
Varying this action with respect to $\chi$, we have $f''(\chi)(R-\chi)=0$.
If $f''(\chi)\neq 0$, we have $R=\chi$.
Substituting this into Eq. \eqref{equiv} leads to Eq. \eqref{fr}.
Redefining the field  by $\theta=f'(\chi)$
and setting $U(\theta)=\theta \chi(\theta)-f(\chi(\theta))$, we have
\begin{equation}\label{jordan}
S={1\over 16\pi G}\int d^4 x\sqrt{-g}\,[\theta R-U(\theta)]+ S_m(g_{\mu\nu},\Psi_m).
\end{equation}
The action \eqref{jordan} is in the Jordan frame, which should be transformed into the Einstein frame to utilize the results of the prior studies \cite{Zhang2016,Zhang2017}, although the chameleon mechanism also works in the Jordan frame \cite{PhysRevD.80.104002}.

Defining the metric in Einstein frame as  $\tilde{g}_{\mu\nu}=\theta g_{\mu\nu}$,
we get the Einstein frame action as follows \cite{RevModPhys.82.451},
\begin{equation}
S_E ={1\over 16\pi G}\int d^4 x\sqrt{-\tilde{g}}[\tilde{R}-\frac{3}{2\theta^2}(\tilde{\partial}\theta)^2-\frac{U(\theta)}{\theta^2}]+S_m(\theta^{-1}\tilde{g}_{\mu\nu},\Psi_m),
\end{equation}
where $(\tilde{\partial}\theta)^2=\tilde{g}^{\mu\nu}\partial_\mu\theta \partial_\nu\theta$ and $\tilde{R}$ is the Ricci scalar of $\tilde{g}_{\mu\nu}$. To change the kinetic term into the standard form, we introduce another scalar field $\phi$ that satisfies the following relation $3(\tilde{\partial}\theta)^2/32\pi G\theta^2=(\tilde{\partial}\phi)^2/2$, that is $\frac{d\phi}{d\theta}=-\sqrt{\frac{3}{16 \pi G}}\frac{1}{\theta}$. Solving this differential equation, we have
$\theta=\exp(-\sqrt{\frac{16\pi G}{3}}\phi)$.
The scalar field $\phi$ can be directly related to the Jordan frame Ricci scalar by
\begin{equation}\label{relation2}
f'(R)=\exp(-\sqrt{\frac{16\pi G}{3}}\phi).
\end{equation}
Therefore, the action in the Einstein frame has the form \cite{RevModPhys.82.451},
\begin{equation}\label{s-t2}
S_E=\int d^4 x \sqrt{-\tilde{g}}[\frac{\tilde{R}}{16\pi G}-\frac12(\tilde{\partial}\phi)^2-V(\phi)]+S_m(A^2(\phi)\tilde{g}_{\mu\nu},\Psi_m),
\end{equation}
where the bare potential is
\begin{equation}
V(\phi)=\frac{f'(R)R-f(R)}{16\pi G f'(R)^2}.
\end{equation}
The conformal coupling function is \cite{RevModPhys.82.451}
\begin{equation}\label{A}
A(\phi)=\frac{1}{\sqrt{f'(R)}}=\exp(\frac{\xi\phi}{M_\text{Pl}})
\end{equation}
with the conformal coupling parameter $\xi=1/\sqrt{6}$.
Variation of $S_E$ with respect to $\tilde{g}_{\mu\nu}$ and $\phi$ gives the field equations
\begin{eqnarray}
\tilde{R}_{\mu\nu}&=&8\pi G [\tilde{S}_{\mu\nu}+\partial_\mu\phi\partial_\nu\phi+V(\phi)\tilde{g}_{\mu\nu}],\label{metric}\\
\tilde{\square}\phi&=&\frac{\text{d} V}{\text{d} \phi}-\frac{\tilde{T}}{A}\frac{\text{d}A}{\text{d}\phi},\label{scalar}
\end{eqnarray}
with
\begin{equation}
\tilde{S}_{\mu\nu}\equiv\tilde{T}_{\mu\nu}-\frac12\tilde{g}_{\mu\nu}\tilde{T},
\end{equation}
where $\tilde{T}_{\mu\nu}\equiv(-2/\sqrt{-\tilde{g}})\delta S_m/\delta\tilde{g}_{\mu\nu}$
is the energy-momentum tensor of matter in the Einstein frame, and $\tilde{\square}\equiv\tilde{g}^{\mu\nu}\tilde{\nabla}_\mu\tilde{\nabla}_\nu$. The covariant derivatives $\tilde{\nabla}_\mu$ obey $\tilde{\nabla}_\mu\tilde{g}_{\alpha\beta}=0$.
The scalar field equation can be rewritten as follows:
\begin{equation}
\tilde{\square}\phi=\frac{\text{d}V_\text{eff}}{\text{d}\phi},
\end{equation}
with the effective potential
\begin{equation}\label{Veff}
V_\text{eff}(\phi)\equiv V(\phi)+\rho[A(\phi)-1].
\end{equation}
Here the matter is assumed to be nonrelativistic, and $\rho\equiv-\tilde{T}/A$ is the conserved energy density in the Einstein frame,
which is independent of $\phi$ \cite{PhysRevLett.93.171104}.

\subsection{Chameleon mechanism}
An important consequence of the conformal coupling function $A(\phi)$ is that  matter will generally feel a fifth force mediated by the scalar field. Since the conformal coupling parameter $\xi$ is of order unity, the fifth force will have a significant impact on the motion of particles \cite{Brax2008}. In order to evade the fifth force constraints, the mass of the field should be sufficiently large in high density environment \cite{PhysRevLett.83.3585}. Since scalar field needs to have cosmological effects to accelerate the expansion of the Universe, on cosmological scales, the magnitude of the scalar mass can be Hubble scale to cause the acceleration of the universe. Thus a mechanism is needed to screen the scalar field  in local environment while let the scalar field accelerate the Universe on large scale \cite{Brax2008,Hu:2007nk}.

The behavior of the scalar field is governed by the effective potential $V_\text{eff}(\phi)$. An essential element of the model is the fact that $V_\text{eff}(\phi)$
depends explicitly on the matter density, as seen in
Eq. \eqref{Veff}. The shape of the effective potential is determined by the function $f(R)$. For a suitably chosen function $f(R)$, the effective potential can have a minimum. We denote by $\phi_\text{min}$ the value at the minimum, that is \cite{Zhang2016},
\begin{equation}
\left.\frac{\text{d}V_\text{eff}}{\text{d}\phi}\right |_{\phi_\text{min}}=0.
\end{equation}
Whilst the mass of small fluctuations around the minimum is \cite{Brax2008},
\begin{equation}\label{mass}
 \left. m^2_\text{eff}=\frac{\text{d}^2V_\text{eff}}{\text{d}\phi^2}\right |_{\phi_\text{min}}=\left[\frac{\text{d}^2V}{\text{d}\phi^2}+\frac{\xi^2}{M_\text{Pl}^2}\rho\exp(\frac{\xi\phi}{M_\text{Pl}})\right ]_{\phi_\text{min}}.
\end{equation}
It can be observed that the scalar field has a density dependent mass. When the density of the environment is large enough, the mass becomes large, and the corresponding fifth force range is so small that it cannot be detected by gravitational experiments \cite{Brax2008}. Laboratory constraints can be greatly alleviated if the mass develops a strong dependence on the ambient density of matter.
Theories in which such a dependence is realized are called to have a chameleon mechanism. Therefore, if the following three conditions can be satisfied in some regions of $\phi$ space, the $f(R)$ model can have a chameleon mechanism \cite{Joyce20151}: (1) $V'(\phi)<0$: The effective potential $V_\text{eff}$ has a minimum; (2) $V''(\phi)>0$: The mass squared $m^2_\text{eff}$ is positive; (3) $V'''(\phi)<0$: The mass can increase with density.

Using Eq. \eqref{relation2}, these conditions can be translated into \cite{Brax2008}
\begin{eqnarray}
V'(\phi)&=&\frac{\xi M_\text{Pl}}{f'^2}[Rf'-2f]<0,\label{chameleon1}\\
V''(\phi)&=&\frac13[\frac{R}{f'}+\frac{1}{f''}-\frac{4f}{f'^2}]>0,\\
V'''(\phi)&=&\frac{2\xi}{3M_\text{Pl}}[\frac{3}{f''}+\frac{f'f'''}{f''^3}+\frac{R}{f'}-\frac{8f}{f'^2}]<0\label{chameleon3}.
\end{eqnarray}

\subsection{Palatini $f(R)$ gravity}

Previous discussions have focused on the metric formalism. We now discuss the Palatini formalism. The action in the Palatini formalism is formally the same as in the metric formalism. However, the Ricci tensor is constructed from the independent connection and is not related to the metric tensor. The Palatini action takes the form \cite{RevModPhys.82.451}
\begin{equation}
S_p={1\over 16\pi G}\int d^4 x\sqrt{-g}\,f(\mathcal R)+ S_m(g_{\mu\nu},\Psi_m).
\end{equation}
Here $\mathcal{R}\equiv g^{\mu\nu}\mathcal{R}_{\mu\nu}$ and the Ricci tensor $\mathcal{R}_{\mu\nu}$ is determined by the independent connection $\Gamma^\mu_{\alpha\beta}$.
Variations with respect to the metric and the connection can yield the following formulae respectively \cite{RevModPhys.82.451},
\begin{equation}
f'(\mathcal{R})\mathcal{R}_{(\mu\nu)}-\frac12 f(\mathcal{R})g_{\mu\nu}= 8\pi G T_{\mu\nu},
\end{equation}
and
\begin{equation}\label{connection}
\nabla_\mu[\sqrt{-g}(\delta^\mu_\alpha f'g^{\beta\nu}-\frac12\delta^\beta_\alpha f'^{\mu\nu}-\frac12 \delta^\nu_\alpha f'g^{\beta\mu})]=0.
\end{equation}

Transforming the action into the Einstein frame, we obtain \cite{RevModPhys.82.451}
\begin{equation}
S_{E}'=\int d^4 x \sqrt{-\tilde{g}}[\frac{\tilde{R}}{16\pi G}-V(\theta)]+S_m(\theta^{-1}\tilde{g}_{\mu\nu},\Psi_m),
\end{equation}
which follows the scalar field equation,
\begin{equation}
2\theta \frac{d V}{d \theta}+\tilde{T}=0.
\end{equation}
Note that, the scalar field $\theta$ is algebraically related to $\tilde{T}$, i.e., $\theta=\theta(\tilde{T})$, which is non-dynamical and cannot propagate in  spacetime. Therefore, we cannot define a mass of the scalar field $\theta$, as discussion above. As a result of the non-dynamical nature of the scalar field, the chameleon mechanism does not apply to Palatini $f(R)$ gravity. There exists another significant difference between Palatini $f(R)$ gravity and the chameleon theory: Since the fifth force is produced by the gradient of the scalar field, and in chameleon theories a compact object in a homogeneous background can generate a scalar field with Yukawa profile. A test particle in the homogeneous background can feel the fifth force. While in Palatini $f(R)$ gravity, the scalar field does not have gradient in a homogeneous background and does not mediate a fifth force. In addition, there are other serious shortcomings of Palatini $f(R)$ gravity \cite{RevModPhys.82.451}.
So, in the rest of this paper, we will only focus on metric $f(R)$ gravity.

\subsection{Stability issues}

More recent attention has focused on the stability issues about metric $f(R)$ gravity.
These include Ostrogradski instability \cite{Woodard2007}, Frolov instability \cite{PhysRevLett.101.061103}, Dolgov-Kawasaki instability \cite{DOLGOV20031} and instability of de Sitter space \cite{PhysRevD.72.124005}. A scrutiny of these issues is needed to make sure that $f(R)$ gravity is viable. The first two stability issues can be bypassed in the specific models discussed below \cite{Woodard2007,PhysRevD.80.064002}.

Dolgov and Kawasaki \cite{DOLGOV20031} found that the Ricci scalar is instable in the $f(R)$ model proposed by Carroll \textit{et al.} \cite{PhysRevD.70.043528}. Their analysis is generalized to a general function by Faraoni \cite{PhysRevD.74.104017}. The origin of this issue is that the mass squared of the scalar degree of freedom is negative. Since the mass squared has the same sign as $f''(R)$, the stability condition can be written as \cite{RevModPhys.82.451}
\begin{equation}
f''(R)>0,\quad \text{for}\; R\geq R_0(>0),
\end{equation}
where $R_0$ is the Ricci scalar today. This condition is satisfied for all the models studied in the following section.

In order to investigate the stability of de Sitter space, we consider a spatial flat Friedmann-Lema\^{\i}tre-Robertson-Walker (FLRW) universe. The vacuum field equations take the form \cite{RevModPhys.82.451}
\begin{eqnarray}
H^2&=&\frac{1}{3 f'}(\frac{R f'}{2}-\frac{f}{2}-3H\dot f'),\nonumber\\
\dot H&=&-\frac{1}{2f'}(\ddot{f'}-H\dot{f'}), \label{frid1}
\end{eqnarray}
where an overdot denotes differentiation with respect to $t$. The stationary points of the dynamical system \eqref{frid1} are de Sitter space with Hubble constant  $H$. The condition for the existence of de Sitter space is \cite{RevModPhys.82.451}
\begin{equation}\label{dS}
Rf'-2f=0.
\end{equation}
The stability condition  of  de Sitter space with respect to inhomogeneous linear perturbations reads \cite{PhysRevD.72.124005}
\begin{equation}\label{dS_stable}
\frac{f'}{f''}-R\geq 0.
\end{equation}
If the solution to Eq. \eqref{dS} meets the stability condition \eqref{dS_stable}, the Universe will enter into a stable de Sitter phase in the future \cite{Xu2014}.

We now impose the stability condition of de Sitter space on the specific $f(R)$ models.
We will investigate the following well studied models
\begin{eqnarray}
(A)\; f(R)&=&R- m^2\frac{c_1(R/m^2)^n}{c_2(R/m^2)^n+1} \;(c_1,c_2,n>0),\label{Hu}\\
(B)\; f(R)&=&R-\mu R_c \tanh\frac{R}{R_c}\;(\mu,R_c>0),\label{Tsu}\\
(C)\; f(R)&=&R-\mu R_c[1-(1+\frac{R^2}{R_c^2})^{-k}]\;(\mu,k,R_c>0).\label{Star}
\end{eqnarray}
The models (A), (B) and (C) are proposed by Hu and Sawicki \cite{Hu:2007nk}, Tsujikawa \cite{PhysRevD.77.023507} and Starobinsky \cite{starobinsky2007disappearing}, respectively.
In the model (A), the mass scale is chosen to be \cite{Hu:2007nk}
\begin{equation}
m^2=\frac{8\pi G \bar{\rho}_0}{3},
\end{equation}
where $\bar{\rho}_0 $ is the average matter density in the universe today.
In the models (B) and (C), $R_c$ roughly corresponds to the order of observed cosmological constant for $\mu=\mathcal{O}(1)$.
During the whole expansion history of the Universe, the Ricci scalar is in the high curvature region, i.e., $R\gg m^2$ or $R\gg R_c$ \cite{Hu:2007nk}. Thus, the model (A) can be approximated by
\begin{equation}\label{Hu:ap}
f(R)=R-\frac{c_1}{c_2}m^2+\frac{c_1}{c_2^2}m^2(\frac{m^2}{R})^n
\end{equation}
and the model (C) can be approximated by
\begin{equation}
f(R)=R-\mu R_c+\mu R_c(\frac{R_c}{R})^{2k}.
\end{equation}
It can be observed that the free parameters of the model (A) are in one-to-one correspondence with that of the model (C) through the relations $m^2c_1/c_2 \rightarrow\mu R_c$, $m^{2(n+1)}c_1/c_2^2\rightarrow\mu R_c ^{2k+1}$ and $n\rightarrow2k$. So we only study the models (A) and (B) in the following.

The model (A) can be expressed as another useful form \cite{PhysRevD.77.023507}
\begin{equation}\label{usefull}
f(R)=R-\alpha R_c \frac{(R/R_c)^n}{(R/R_c)^n+1},
\end{equation}
where  $\alpha=c_1c_2^{{1}/{n}-1}$ and $R_c=m^2c_2^{-{1}/{n}}$. The following relation holds at the de~Sitter point: \cite{de2010f}
\begin{equation}\label{alpha}
\alpha=\frac{(1+x^n)^2}{x^{n-1}(2+2x^n-n)},
\end{equation}
where $x\equiv R/R_c$. The stability condition \eqref{dS_stable} implies the relation \cite{de2010f},
\begin{equation}
2x^{2n}-(n-1)(n+4)x^n+(n-1)(n-2)\geq 0.
\end{equation}
Thus for each specific $n$, the above inequality gives a bound on $x$ and this bound can be transformed into a bound on $\alpha$ through Eq. \eqref{alpha}. For instance, when $n=2$, one has $x\geq\sqrt{3}$ and $\alpha\geq 8\sqrt{3}/9$.
In the following section, we will come back to discuss this inequality.

\section{Constraints on $f(R)$ gravity }\label{constraint}

In this section, we consider the observational constraints on metric $f(R)$ gravity in cosmological scale, solar system and binary pulsar systems, respectively.

\subsection{Cosmological constraints}\label{cosmos_const}

In order to satisfy the tests on cosmological scales, the $f(R)$ models should mimic the $\Lambda$CDM model at the late time and provide an effective cosmological constant. Similar to the previous work \cite{Zhang2016}, in this paper we do not consider the cosmological perturbations of $f(R)$ gravity \cite{de2010f}. We leave this issue for the general $f(R)$ gravity as a future work. The bare potential $V(\phi)$ in action \eqref{s-t2} can provide the effective cosmological constant to accelerate the universe expansion, which is given by \cite{Zhang2016}
\begin{equation}\label{cosmos}
\Lambda_\text{eff}=8\pi G V_\text{VEV}=\left. \frac{R f'(R)-f(R)}{2 f'(R)^2}\right |_{R=R_\infty},
\end{equation}
where $R_\infty$ is the background value of Ricci scalar.
In order to mimic the $\Lambda$CDM model, we need that the value of $\Lambda_\text{eff}$ is equal to the observed cosmological constant $\Lambda$, which accelerates the cosmic expansion.

Now, we can apply the cosmological constraint \eqref{cosmos} to specific
$f(R)$ models. For model  (A), we substitute  Eq. \eqref{Hu:ap} into Eq. \eqref{cosmos}, and obtain
\begin{equation}
\Lambda_\text{eff} \approx \frac{c_1}{2c_2} m^2,
\end{equation}
which, in turn, implies that
\begin{equation}\label{Hu:cosmos}
\frac{c_1}{c_2}\approx \frac{2\Lambda_\text{eff}}{m^2}=6 \frac{\Omega_\Lambda}{\Omega_m}=13.5.
\end{equation}
Note that, we adopted the density parameters  $\Omega_m=0.308$ and $\Omega_\Lambda=0.692$  \cite{ade2016planck}.
This expression of $c_1/c_2$ is consistent with Eq. (26) in \cite{Hu:2007nk}. Now it can be seen from Eq. \eqref{Hu:ap} that there are two remaining parameters $n$ and
$c_1/c_2^2$  in this model.

 Using the relation $\alpha=c_1c_2^{{1}/{n}-1}$ and the cosmological constraint \eqref{Hu:cosmos}, we have
\begin{equation}
\frac{c_1}{c_2^2}=13.5(\frac{13.5}{\alpha})^n.
\end{equation}
Thus, in the case $n=2$, the stability condition $\alpha\geq 8\sqrt{3}/9$ implies an upper bound on $c_1/c^2_2$
\begin{equation}
\frac{c_1}{c_2^2}\leq 1038.
\end{equation}
Using Eq. \eqref{Hu:ap}, we have
\begin{equation}
\frac{c_1}{c_2^2}=\frac{1-f'(R_0)}{n}\left(\frac{R_0}{m^2}\right)^{n+1}.
\end{equation}
For a spatial flat FLRW universe, the scalar curvature at the present epoch is $R_0=m^2(12/ {\Omega_m}-9)$ \cite{Hu:2007nk}.
Consequently, for different $n$, we can obtain different upper bounds on $|f'(R_0)~-~1|$.
The results are presented in Fig. \ref{huc1c2} with dotted line.

Similarly, for the model (B), the cosmological constraint is
\begin{equation} \label{Tsu:cosmos}
\Lambda_\text{eff} \approx \frac{\mu R_c}{2},
\end{equation}
and the stability condition \eqref{dS_stable} implies that \cite{de2010f}
\begin{equation}\label{stableB}
\mu >0.905.
\end{equation}

\subsection{Solar system constraints}\label{local}
In the solar system, the gravitational field is weak and the velocity of planets is slow compared with the speed of light. Thus we can apply the PPN formalism to solar system tests. In the PN limit, the spacetime metric predicted by different metric theory of gravity has the same structure and can be characterized by ten PPN parameters \cite{Will2014}. Among them, the most important parameters are $\gamma$ and $\beta$.

Here, we derive the PPN parameters  $\gamma$ and $\beta$ and the effective gravitational constant $G_\text{eff}$ in the general metric $f(R)$ gravity with chameleon mechanism. For a scalar-tensor theory with action \eqref{s-t2} , the solution to the scalar field equation \eqref{scalar} is given by \cite{Zhang2017}
\begin{equation}
\phi(r)=\phi_\infty -\epsilon M_\text{Pl}\frac{G M_E}{r}e^{-m_\infty r},
\end{equation}
where the screened parameter is defined as,
\begin{equation}
\epsilon\equiv\frac{\phi_\infty-\phi_0}{M_\text{Pl}\Phi_E}.
\end{equation}
The parameters $M_\text{E}$ and $\Phi_\text{E}\equiv G M_\text{E}/r$ are the mass and the Newtonian potential at the surface of the source object in the Einstein frame, respectively. The quantity
$\phi_0$ is the field in side the source object and $\phi_\infty$ is the field in the background environment. $m_{\infty}$ is the effective mass of scalar field at $\phi=\phi_{\infty}$.

In order to solve the metric field equations, we make use of the PPN formalism introduced in \cite{Will2014,will1993theory}. In this formalism, the gravitational field of the source is weak $GM/r\ll 1$, and the typical velocity $\vec{v}$ of the source is small, i.e. $v^2\sim GM/r\ll 1$. Thus, we can use the perturbative expansion method to solve the field equations, and all dynamical quantities can be expanded to $\mathcal{O}(n)\propto v^{2n}$. The metric field $g_{\mu\nu}$ can be expanded around the Minkowski background as follows:
\begin{equation}
g_{\mu\nu}=\eta_{\mu\nu}+\accentset{(1)}h_{\mu\nu}+\accentset{(2)}h_{\mu\nu}+ \mathcal{O}(3).
\end{equation}

We solve the field equations \eqref{metric} and \eqref{scalar} using the PPN method \cite{will1993theory}, and transform the metric to the Jordan frame. Making use of the definitions of $\gamma$ and $\beta$ as follows \cite{Zhang2016},
\begin{eqnarray}
\accentset{(1)}h_{\text J 00}=\frac{2G_\text{eff}M_\text J}{\chi},~~~
\accentset{(1)}h_{\text J \chi\chi}=\gamma \frac{2G_\text{eff}M_\text J}{\chi},~~~
\accentset{(2)}h_{\text J 00}=-\beta \frac{4G^2_\text{eff}M^2_\text J}{2\chi^2},
\end{eqnarray}
where $M_J$ and $\chi$ are the mass and radial coordinate in the Jordan frame, respectively.
We obtain the PPN parameters
\begin{eqnarray} \label{ppn}
\gamma=1-\frac{2A_1}{A_\text{VEV}}M_\text{Pl}\epsilon,~~
\beta=1-M_\text{Pl}^2(\frac{A^2_1}{2A^2_\text{VEV}}-\frac{A_2}{A_\text{VEV}})\epsilon^2,~~
G_\text{eff}=G A^2_\text{VEV}(1+\frac{A_1}{A_\text{VEV}}M_\text{Pl}\epsilon),\label{G}
\end{eqnarray}
where $A_\text{VEV}$,$A_1$ and $A_2$ are the expansion coefficients of $A(\phi)$, i.e.,
\begin{equation}
A(\phi)=A_\text{VEV}+A_1(\phi-\phi_\infty)+A_2(\phi-\phi_\infty)^2+\cdots.
\end{equation}
Note that, here we have taken the limit $m_\infty r\ll 1$, since in the solar system, the distance $r$ is always much less than the Compton wavelength $m^{-1}_\infty$ \cite{Zhang2016}.

Applying to the general metric  $f(R)$ gravity, using Eqs. \eqref{relation2} and \eqref{A} we obtain the expansion coefficients
\begin{eqnarray}
A_\text{VEV}=\frac{1}{\sqrt{f'(R_\infty)}}, ~~~ A_1=\frac1{\sqrt{6}M_\text{Pl}}\frac{1}{\sqrt{f'(R_\infty)}},~~~A_2=\frac1{12M^2_\text{Pl}}\frac{1}{\sqrt{f'(R_\infty)}}.
\end{eqnarray}
Following the discussion of \cite{Hu:2007nk}, $R_\infty=8\pi G \rho_g$ and $\rho_g=10^{-24}\text{g\,cm}^{-3}$ is the average galactic density in the solar vicinity. In the solar system, the source object of the scalar field is the Sun and the background is the Milky Way.
Since the density of the Sun is much higher compared with the galactic background, we have $\phi_\infty\gg\phi_0$. Then the screened parameter can be approximated by
\begin{equation}
\epsilon=\frac{\phi_\infty}{M_\text{Pl}\Phi_\text{E}}.
\end{equation}

Substituting the above parameters into Eq. \eqref{G}, we obtain
\begin{eqnarray}
\gamma=1+\frac{\ln f'(R_\infty)}{\Phi_\text{E}},~~~
\beta=1,~~~
G_\text{eff}=\frac{G}{f'(R_\infty)}(1-\frac{\ln f'(R_\infty)}{2\Phi_\text{E}}).\label{Geff}
\end{eqnarray}
We find that the expression of parameter $\gamma$ is consistent with Eq. (64) in \cite{Hu:2007nk}. The parameter $\beta$ is unity no matter whatever the functional form $f(R)$ is. As can be seen from Eq. \eqref{G}, when the conformal coupling function $A(\phi)$ has the exponential form, the two terms in the bracket of the expression of $\beta$ cancel each other out. And Eq. \eqref{A} shows that the conformal coupling function $A(\phi)$ always has the exponential form in metric $f(R)$ gravity. This suggests that the experimental tests of parameter $\beta$ cannot distinguish between GR and metric $f(R)$ gravity. The relation between $\gamma$ and $G_\text{eff}$ is
\begin{equation}
\frac{G_\text{eff}}{G}-1\approx-\frac{\gamma-1}{2}.
\end{equation}

Using the Cassini constraint  $|\gamma-1|<2.3\times 10^{-5}$ \cite{bertotti2003test} and the  Newtonian potential at the surface of the Sun $\Phi_\text{E}=2.12\times 10^{-6}$, we obtain the constraint on general $f(R)$ gravity as follows,
\begin{equation}
|\ln f'(R_\infty)|=|\gamma-1|\Phi_\text{E}<4.9\times10^{-11}.
\end{equation}
Since $\ln f'(R_\infty)\approx f'(R_\infty)-1$, we have
\begin{equation}\label{solar}
|f'(R_\infty)-1|<4.9\times10^{-11}.
\end{equation}
Note that, this is a general constraint for any metric $f(R)$ gravity with chameleon mechanism, which is independent of the form of $f(R)$.

We can apply this solar system constraint to the models (A) and (B). In the model (A), using Eq. \eqref{Hu:ap}, we have \cite{Hu:2007nk}
\begin{equation}
\Big(\frac{1-f'(R_\infty)}{1-f'(R_0)}\Big)^{\frac{1}{n+1}}=\frac{R_0}{8\pi G \rho_g}=8.14\times 10^{-7} \frac{R_0}{m^2}\frac{\Omega_m h^2}{0.13}\Big(\frac{\rho_g}{10^{-24}~\text{g cm}^{-3}}\Big)^{-1}
\end{equation}
Here, $R_\infty=8\pi G \rho_g$ and $\rho_g=10^{-24}\text{g\,cm}^{-3}$ is the average galactic density in the solar vicinity. We adopted  the physical matter density $\Omega_m h^2=0.1415$ \cite{ade2016planck}. 

\begin{figure}
\begin{center}
\includegraphics[width=8cm, height=6.2cm]{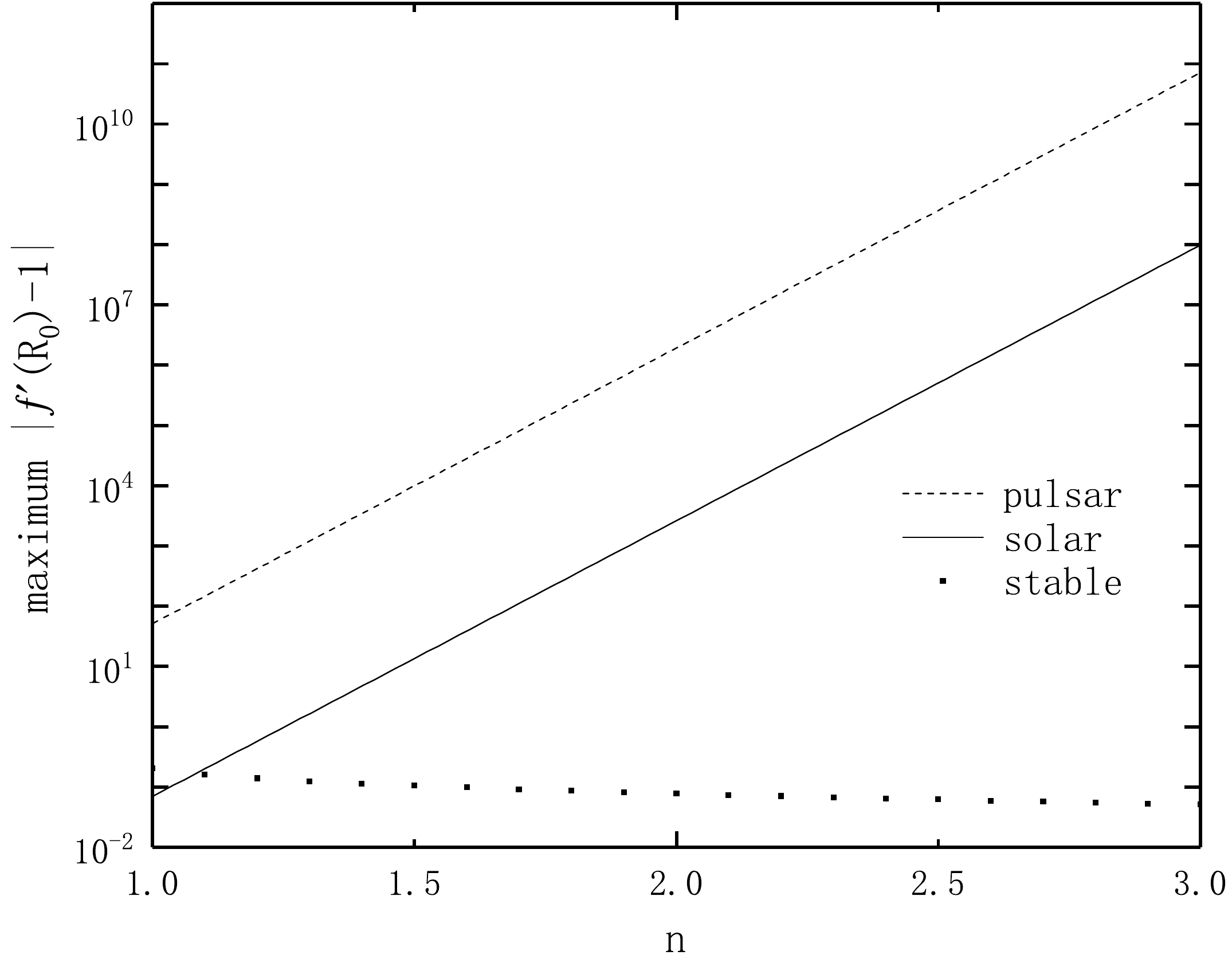}
\caption{Maximum value of $|f'(R_0)-1|$ in Hu-Sawicki model allowed by the solar system constraint (solid line) , the pulsar constraint (dashed line) and the stability condition (dotted line), respectively. Note that, for the dotted line, we have considered the cosmological constraint.}\label{huc1c2}
\end{center}
\end{figure}

Applying inequality \eqref{solar} to the above equation, we have
\begin{equation}
|f'(R_0)-1|<4.9\times 10^{-11}\Big(\frac{8\pi G \rho_g}{R_0}\Big)^{n+1}.
\end{equation}
The equivalence principle places a bound on the parameter of model (A)  \cite{PhysRevD.77.107501}
\begin{equation}
n>1.8.
\end{equation}
As shown in Fig. \ref{huc1c2}, in the region $1.8<n<3$, the solar system constraint (solid line)  is  fairly weak when compared with the stability condition (dotted line) and  is  sensitive to the value of $n$.

Similarly, in the model (B) we have
\begin{equation}
1-f'(R_\infty)=\frac{\mu}{\cosh^2\frac{\mu R_\infty}{2\Lambda_\text{eff}}},
\end{equation}
where Eq. \eqref{Tsu:cosmos} was used to eliminate $R_c$ in terms of $\mu$. The solar system constraint \eqref{solar} yields
\begin{equation}\label{solarB}
\mu > 9.5\times 10^{-5}.
\end{equation}
Compared with the stability condition \eqref{stableB}, this shows that the solar system constraint on $f(R)$ gravity is weaker for this model .

Assuming that the cosmological constraint \eqref{cosmos}  and solar system constraint \eqref{solar} are both satisfied, we have checked that in the model (A) and (B), the conditions for chameleon mechanism  \eqref{chameleon1}-\eqref{chameleon3} can all be satisfied.

\subsection{Binary pulsar constraints}\label{pulsar_const}

It is well known that the compact binary systems can lose the orbital energy due to gravitational radiation, and the orbital period will decay. In different theories of gravity, the decay rates are different \cite{will1993theory,Will2014}, which provides another independent opportunities to test the metric $f(R)$ gravity. In a binary system, when the difference between the screened parameters of the two compact stars is significant, the dipole radiation dominates the orbital decay rate. Since the screened parameter is inversely proportional to surface gravitational potential, the neutron star-white dwarf (NS-WD) systems are the best testbeds to constrain the parameters of  $f(R)$ gravity.  In the previous work \cite{Zhang2017}, we have studied this effect in the most general scalar-tensor gravity with screening mechanism. For a quasicircular ($e\ll 1$) NS-WD  binary system,
the orbital period derivative is given by \cite{Zhang2017}
\begin{equation}\label{pdot}
\dot{P}=\dot{P}^{\rm GR}\left[1+\frac{5}{192}\Big(\frac{P}{2\pi Gm}\Big)^{2/3}(\epsilon_{\rm WD}-\epsilon_{\rm NS})^2\right]\,\,\,.
\end{equation}
Here, $P$ denotes the orbital period, $m=m_{\rm NS}+m_{\rm WD}$ is the total mass, $\mu=m_{\rm NS}m_{\rm WD}/m$ is the reduced mass, $\epsilon_{\rm WD}={\phi_\infty}/{M_{\rm Pl} \Phi_{\rm WD}}$ and $\epsilon_{\rm NS}={\phi_\infty}/{M_{\rm Pl} \Phi_{\rm NS}}$  are the screened parameter of the white dwarf and the neutron star respectively and
\begin{equation}
\dot{P}^{\rm GR}=-\frac{192\pi}{5}\left(\frac{2\pi Gm}{P}\right)^{5/3}\!\!\left(\frac{\mu}{m}\right)
\end{equation}
represents the GR prediction of the orbital period derivative. The second term in Eq. \eqref{pdot} corresponds to the scalar dipole radiation correction.

We apply this result to the general metric $f(R)$ gravity with chameleon mechanism. Using Eq. \eqref{relation2}, the orbital period derivative translates into
\begin{equation}
\frac{\dot{P}}{\dot{P}^{\rm GR}}=1+\frac{15}{384}\Big(\frac{P}{2\pi Gm}\Big)^{2/3}\Big(\frac{\ln f'(R_\infty)}{\Phi_{\rm WD}}-\frac{\ln f'(R_\infty)}{\Phi_{\rm NS}}\Big)^2 \, .
\end{equation}
It can be seen that in the special case $f(R)=R-2\Lambda$, the above result is  reduced to $\dot{P}=\dot{P}^{\rm GR}$. Because $\Phi_{\rm NS}/\Phi_{\rm WD}\sim 10^4$, the orbital period derivative can be approximated by
\begin{equation}\label{Aobs}
\frac{\dot{P}}{\dot{P}^{\rm GR}}=1+\frac{15}{384}\Big(\frac{P}{2\pi Gm}\Big)^{2/3}\Big(\frac{\ln f'(R_\infty)}{\Phi_{\rm WD}}\Big)^2 \, .
\end{equation}
Since all the pulsar observation agrees well with the GR prediction within the errors \cite{Stairs2003,Antoniadis1233232,freire2012relativistic}, the observation value of the period derivative can be expressed as
\begin{equation}
\frac{\dot{P}^{\rm obs}}{\dot{P}^{\rm GR}}=1+\delta\pm\sigma
\end{equation}
where $\delta$ is the fractional deviation of the observed $\dot{P}^{\rm obs}$ from the GR prediction, $\sigma$ is the observational uncertainty.
Thus the background field value $f'(R_\infty)$ cannot deviate from unity too much, that is,
\begin{equation}\label{psr_bound}
|\ln f'(R_\infty)|\approx|f'(R_\infty)-1|<(|\delta|+2\sigma)^{\frac12}(\frac{m}{M_\odot})^{\frac13}(\frac{P}{1 \text{d}})^{-\frac13}(\frac{m_{\rm WD}}{M_\odot})(\frac{R_{\rm WD}}{R_\odot})^{-1}\times 7.63\times 10^{-9}
\end{equation}
at 95\% confidence level.

Up to now, more than 2500 pulsars have been observed \cite{neutron}. However, most of them are isolated and their mass cannot be determined. Table 2 in  \cite{neutron} lists fifteen NS-WD systems with low-eccentricity orbits which have accurate  measurement of mass. Among these fifteen NS-WD systems only PSR J0348~+0432 and PSR J1738~+0333 have accurate observation value of the radius of the white dwarf companion. Thus we use these two NS-WD systems to constrain $f(R)$ gravity and  list here the relevant parameters  in Table \ref{psr}.

In the PSR J0348 +0432 case (see Table \ref{psr}), $\delta=0.05$ and $\sigma=0.18$. Substituting the parameters into inequality \eqref{psr_bound}, we obtain the upper bound 
\begin{equation}
|f'(R_\infty)-1|<3.583\times10^{-8}
\end{equation}
at 95\% confidence level. Similarly, using the observation data of PSR J1738~+0333, we obtain
\begin{equation}
|f'(R_\infty)-1|<3.579\times10^{-8}
\end{equation}
at 95\% confidence level.
Compared with the solar system constraint \eqref{solar}, the pulsar constraint is three orders of magnitude weaker.

Applying the pulsar constraint to the model (A), we obtain
\begin{equation}
|f'(R_0)-1|<3.6\times 10^{-8}\Big(\frac{8\pi G \rho_g}{R_0}\Big)^{n+1}.
\end{equation}
The above result is also shown in Fig. \ref{huc1c2} with dashed line. Similarly, applying the pulsar constraint to the model (B), we obtain
\begin{equation}
\mu > 5.4\times 10^{-5}.
\end{equation}
Consistently, we find both of them are relatively weaker than the corresponding constraints of solar system.

\begin{center}
\begin{table*}[htb]
\caption{Parameters of the binary systems  with 1-$\sigma$  uncertainties.}
\label{psr}
\begin{tabular}{l r r}
\hline
\hline
PSR & J0348 +0432  \cite{Antoniadis1233232} & J1738 +0333 \cite{freire2012relativistic} \\
\hline
Eccentricity, $e$  & $\sim10^{-6}$ & $(3.4\pm1.1) \times 10^{-7}$\\
Period, $P$ (day) & 0.102424062722(7) &0.3547907398724(13) \\
Period derivative, $\dot{P}$ ($10^{-14}$) & $-27.3\pm 4.5$ & $-2.59\pm0.32$\\
$\dot{P}^{\rm obs}/\dot{P}^{\rm GR}$ & $1.05\pm0.18$ & $0.93\pm0.13$\\
Total mass, $m$ ($M_\odot$) & $2.18\pm0.04$ & $1.65_{-0.06}^{+0.07}$ \\
WD mass, $m_{\rm WD}$ ($M_\odot$) & $0.172\pm0.003$ & $0.181_{-0.007}^{+0.008}$ \\
WD radius, $R_{\rm WD}$ ($R_\odot$) & $0.065\pm0.005$ & $0.037_{-0.003}^{+0.004}$\\

\hline
\hline
\end{tabular}
\end{table*}
\end{center}

\section{Conclusions}\label{conclusion}
The $f(R)$ gravity has been extensively studied to explain the accelerating expansion of the universe. In this paper, we have studied the general $f(R)$ gravity through the scalar-tensor representation. In this theory, the chameleon mechanism is crucial for $f(R)$ gravity to escape the fifth force constraints. However, due to the non-dynamical nature of the scalar field in Palatini $f(R)$ gravity, this mechanism does not apply to the theory. Therefore, we focused on the metric $f(R)$ gravity with chameleon mechanism.

We calculated the PPN parameters $\gamma$ and $\beta$ for the general $f(R)$ gravity, and found that $\beta=1$ in the limit $m_\infty r\ll 1$. As a result, the observed value of $\beta$ cannot constrain the parameters of $f(R)$ models. Applying  the Cassini spacecraft measurement of $\gamma$, we obtained the constraint $|f'(R_\infty)-1|<4.9\times10^{-11}$ on the metric $f(R)$ gravity, which is consistent with the previous works. To pass the cosmological test, the metric $f(R)$ gravity should provide the effective cosmological constant. We also calculated the effective cosmological constant in $f(R)$ gravity. In general, the cosmological constraint can reduce one free model parameter in a given specific $f(R)$ model.

In addition, we calculated  the orbital period derivative $\dot{P}$ of binary pulsar systems in the metric $f(R)$ gravity. Since GR has survived the binary pulsar test, the $\dot{P}$ in the metric $f(R)$ gravity cannot deviate from that in GR too much. We found that  the pulsar constraint from the observations of  PSR J0348~+0432 and PSR J1738~+0333 is $|f'(R_\infty)-1|<3.6\times10^{-8}$. This is relatively weaker than the current constraints derived from the solar system observations. We also studied the stability condition of de Sitter space. Compared with the observational constraints (binary pulsar and solar system), this theoretical constraint is more stringent in Hu-Sawicki model and Tsujikawa model. With the chameleon mechanism, the metric $f(R)$ gravity with suitable parameters can pass the cosmological test, the solar system test and the binary pulsar test at the same time.

\section*{Acknowledgements}
This work is supported by NSFC Grants Nos. 11773028, 11603020, 11633001, 11173021, 11322324, 11653002 and 11421303, the project of Knowledge Innovation Program of Chinese Academy of Science, the Fundamental Research Funds for the Central Universities and the Strategic Priority Research Program of the Chinese Academy of Sciences Grant No. XDB23010200.

\end{document}